\documentclass[trackchanges]{aastex701}

\usepackage{threeparttable} 
\usepackage{booktabs}
\usepackage{multirow}
\usepackage{longtable}
\usepackage{placeins}
\usepackage{afterpage}
\usepackage{fancyhdr}
\usepackage{amsmath}
\usepackage{ulem}
\usepackage{xcolor}

\shorttitle{Astrometry of Phobos}
\shortauthors{Lai et al.}

\begin{document}

\title{Automated Shape-Model-Based Astrometry of Phobos from Mars Express SRC Images}

\author{Wangxin Lai}
\affiliation{
Department of Computer Science, Jinan University,
Guangzhou 510632, China
}
\email{xswl233hhh@stu2023.jnu.edu.cn}

\author[0000-0003-4086-9678]{Qingfeng Zhang}
\affiliation{
Department of Computer Science, Jinan University,
Guangzhou 510632, China
}
\affiliation{
Sino-French Joint Laboratory for Astrometry, Dynamics and Space Science, Jinan University, Guangzhou 510632, China
}
\email{tqfz@jnu.edu.cn}

\author{Rui Zhang}
\affiliation{
Department of Computer Science, Jinan University,
Guangzhou 510632, China
}
\email{ruizhang@stu2025.jnu.edu.cn}

\author{Kai Tang}
\affiliation{
Shanghai Astronomical Observatory, Chinese Academy of Sciences, Shanghai 200030, China
}
\email{tangkai@shao.ac.cn}

\author{Chunyu Ding}
\affiliation{
Institute for Advanced Study, Shenzhen University, Shenzhen, 518060, China.
}
\email{dingchunyu@szu.edu.cn}

\author{Zhan Li}
\affiliation{
Department of Computer Science, Jinan University,
Guangzhou 510632, China
}
\affiliation{
Sino-French Joint Laboratory for Astrometry, Dynamics and Space Science, Jinan University, Guangzhou 510632, China
}
\email{lizhan@jnu.edu.cn}

\author{Haipeng Wang}
\affiliation{
School of Computer Science, Northwestern Polytechnical University, Xi’an 710072, China
}
\email{haipeng@nwpu.edu.cn}

\correspondingauthor{Qingfeng Zhang}
\email{tqfz@jnu.edu.cn}

\begin{abstract}
High-resolution spacecraft images provide important astrometric constraints for orbit refinement, but measurements of resolved bodies are often limited by labor-intensive control-point selection and the difficulty of achieving consistent reductions over large image archives. We present an automated shape-model-based astrometric pipeline for Phobos and apply it to Mars Express Super Resolution Channel (SRC) images. For each exposure, a synthetic image is rendered from a high-resolution 3D shape model under the nominal spacecraft-target-Sun geometry. Feature correspondences between the observed and synthetic images are established using SuperPoint and SuperGlue, followed by RANSAC filtering. The matched synthetic-image keypoints are then associated with surface points through ray-shape intersection. The geometric adjustment fixes the adopted body orientation, spacecraft state, and corrected camera pointing and estimates only two effective plane-of-sky position offsets using the exact perspective-projection model. These offsets are used to derive the center-of-figure position of \textit{Phobos}. We first test the method on an image set previously analysed with a control-point approach and obtain comparable astrometric performance. We then extend the analysis to a larger SRC dataset spanning 2007--2025 and obtain 1113 successful measurements. Relative to the JPL MAR099 ephemeris, the resulting observed-minus-computed residuals have mean values of 0.186~km in $\alpha\cos\delta$ and 0.053~km in $\delta$, with corresponding standard deviations of 0.609~km and 0.583~km. These results demonstrate that the proposed pipeline provides a practical approach to large-scale, homogeneous astrometric reduction of archival spacecraft images of \textit{Phobos}, with potential
application to other resolved bodies.
\end{abstract}



\section{Introduction}
Astrometric measurements of resolved bodies are crucial for refining dynamical models, improving ephemerides, and enabling reliable physical interpretation of planetary observations. When a target is resolved and its apparent disk spans many pixels, the measurement task differs fundamentally from point-source astrometry: the object center must be estimated from disk-resolved, illumination-dependent structure, with geometric constraints distributed across the projected disk rather than concentrated in a single point-like centroid.

Over the past decades, disk-resolved astrometry has largely relied on two families of geometry-driven techniques: limb fitting and control-point (landmark) networks. Limb-fitting methods estimate the body center by extracting limb and/or terminator points from the observed image and matching them to the projected outline of a shape model under the viewing and illumination geometry \citep[e.g.,][]{Oberst2006, Tajeddine2013, Pasewaldt2015, Cooper2018}. In practice, the accuracy of limb-fitting methods depends strongly on reliable boundary extraction. Moreover, the resulting astrometric constraints are confined to one-dimensional contours and do not use the interior texture of the illuminated disk. These limitations become particularly severe when the boundary is weakly contrasted or blurred, or when only part of the limb is illuminated and detectable. In addition, viewing- and illumination-dependent errors in the extracted boundary can introduce systematic biases into the fitted astrometric position \citep[e.g.,][]{Cooper2014}. These limitations motivate methods that exploit information distributed across the disk and reduce reliance on boundary extraction.

Control-point approaches, by contrast, associate the image locations of surface landmarks distributed across the illuminated disk with corresponding points on a 3D shape model \citep[e.g.,][]{Willner2010,Burmeister2018}. By using interior surface features, these approaches provide more spatially distributed geometric constraints than limb-fitting methods. In classical workflows, however, landmark localization, correspondence establishment, and match validation often require substantial manual or semi-manual effort. This limits scalability and makes the homogeneous reduction of large archival datasets difficult. The key challenge is therefore to retain disk-wide, shape-model-based constraints while automating correspondence establishment and robust outlier rejection.

Recent progress in computer vision provides precisely such capabilities. Classical local-feature pipelines based on hand-crafted descriptors such as the Scale-Invariant Feature Transform (SIFT; \citealt{Lowe2004}), Speeded-Up Robust Features (SURF; \citealt{Bay2008}) and Oriented FAST and Rotated BRIEF (ORB; \citealt{Rublee2011}) support automatic correspondence generation, while more recent learned matchers such as SuperGlue (\citealt{Sarlin2020}) and LoFTR (\citealt{Sun2021}) improve robustness under substantial appearance and viewpoint changes. These advances make it feasible to establish disk-wide constraints by matching an observed image directly to a synthetic rendering, without manual landmarking.

Phobos provides a compelling target for image-based astrometry. As the larger and innermost moon of Mars, it is an important object for studies of Martian system dynamics and for the planning of current and future exploration missions. Mars Express \citep{Chicarro2004} has provided a long time baseline of disk-resolved Phobos imaging with the Super Resolution Channel (SRC) of the High Resolution Stereo Camera (HRSC) \citep{Jaumann2007,Willner2008}, underpinning major astrometric datasets and ephemeris refinement efforts \citep{Oberst2006,Pasewaldt2015,Ziese2018,Lainey2021}.  Accurate image-based astrometry also depends on the availability of an adequate 3D shape model. Recent high-resolution stereophotoclinometry products \citep{Ernst2023} and related analyses of Phobos shape models \citep{Chen2024} provide a solid geometric basis for synthetic-image projection and image matching. Together, these factors make Phobos a well-suited case for developing more automated, accurate, and scalable image-based astrometric pipelines.

In this work, we present an astrometric pipeline for disk-resolved images of Phobos. After an initial verification of the pointing for each image sequence, the subsequent measurements are performed automatically. For each exposure, computer vision features are matched between the observed image and a synthetic view rendered from a 3D shape model under the nominal observing geometry. The resulting inlier correspondences provide disk-wide constraints for a two-parameter adjustment of the Phobos center-of-figure (COF) position in the plane of the sky. By combining shape-model rendering with automatic feature matching and robust outlier rejection, the method retains the disk-wide geometric constraints of  control-point approaches without requiring manual landmark selection during the astrometric reduction or relying solely on boundary information. It should also be applicable to other disk-resolved bodies, provided that sufficiently accurate shape models and synthetic renderings are available.

The remainder of the paper is organized as follows. Section~\ref{sec:method} describes the automated pipeline, Section~\ref{sec:valdata} presents the data and ancillary products, Section~\ref{sec:results} reports the astrometric results, Section~\ref{sec:discussion} discusses the limitations and implications of the method, and  Section~\ref{sec:conclusion} summarizes the main conclusions.

\section{Method}
\label{sec:method}
The objective of the pipeline is to estimate, for each disk-resolved SRC image of \textit{Phobos}, the COF position that is most consistent with the adopted shape model and imaging geometry. 

Starting from a nominal observation geometry computed using kernel files from the Navigation and Ancillary Information Facility (NAIF) SPICE system\citep{Acton1996,ActonEtAl2018}, the procedure first applies a sequence-level pointing correction and then performs four main operations for each image: synthetic-image generation, feature matching between the observed and synthetic images, approximate 2D--3D association on the shape model, and per-image geometric adjustment based on the retained correspondences. The refined geometry is finally used to determine the COF position and the corresponding astrometric measurement, while images that do not provide sufficiently reliable constraints are rejected through quality control.

\subsection{Per-sequence pointing correction}
\label{sec:pointing}
Before the per-image refinement described below, a preliminary pointing correction was applied to each SRC observing sequence, because the nominal camera pointing is generally not sufficiently accurate for precise astrometric reduction. We adopted the sequence-level correction and interpolation strategy used in previous MEX/SRC reductions \citep{Willner2008,Lainey2021}.

For each observing sequence, star-based pointing corrections were determined
for the first and last long-exposure frames, which served as anchors for the sequence-level interpolation. The expected apparent directions of the reference stars at the time and location of each observation were computed from Gaia DR3 data \citep{vallenari2023gaia}. Gaia source positions, expressed in the International Celestial Reference System (ICRS) at the reference epoch J2016.0, were propagated to each observation epoch using the catalog proper motions. The propagated barycentric directions were converted to apparent directions as seen from Mars Express by applying stellar parallax corrections based on the Gaia parallaxes and the SPICE-derived Mars Express position relative to the solar system barycentre, and by correcting for stellar aberration using the SPICE-derived barycentric velocity of Mars Express. The Mars Express ephemeris and attitude were obtained from SPICE kernels, and the camera-pointing calculations were performed with respect to the SPICE J2000 inertial frame. At the accuracy required for this analysis, the ICRS axes of the Gaia catalog were treated as consistent with the SPICE J2000 axes. The resulting apparent stellar directions were projected onto the image plane using the nominal camera pointing.  Bright stars were identified in each
long-exposure frame, and matching their measured image positions to the
projected catalog positions yielded offsets in the sample and line directions.
These offsets were then used to refine the pointing of the frame.

The pointings of the intermediate \textit{Phobos} images were obtained by
linear interpolation between the corrected pointings of the two anchor frames. Only a small number of usable reference stars were available in these long-exposure frames, typically one or two per frame, with a minimum of one and a maximum of eight. Under these sparse constraints, the correction mainly captured translational offsets in sample and line rather than a fully determined camera-attitude solution. We therefore did not solve for image rotation about the boresight, consistent with earlier MEX/SRC reductions under similarly limited stellar constraints \citep{Willner2008}. Residual errors caused by spacecraft jitter or nonlinear pointing variations \citep{Lainey2021} are regarded as part of the overall astrometric uncertainty. 

\subsection{Synthetic image generation}
\label{sec:render}
Using the corrected pointing, the image-projection model, the observation geometry, and the adopted 3D shape model of \textit{Phobos} \citep{Ernst2023}, represented as a triangular mesh, we generated a synthetic image $I_{\rm syn}$ for subsequent matching. The spacecraft position and camera pointing were evaluated at the mid-exposure time $t_{\rm obs}$. The apparent position vector from Mars Express to the center of \textit{Phobos}, expressed in the SPICE J2000 frame, was computed using the SPICE \texttt{spkpos} routine with the \texttt{LT+S} correction, which accounts for one-way light time and stellar aberration. The returned light time $\tau$ was used to evaluate the orientation of \textit{Phobos} at the target epoch $t_{\rm obs}-\tau$. Each mesh vertex was then transformed from the body-fixed frame to the J2000 frame using this orientation and added to the corrected center vector before image projection. No separate light-time or stellar-aberration correction was applied to individual vertices; differential effects across the surface of \textit{Phobos} were neglected.

\begin{figure}[h]
    \centering
    \includegraphics[width=1\linewidth]{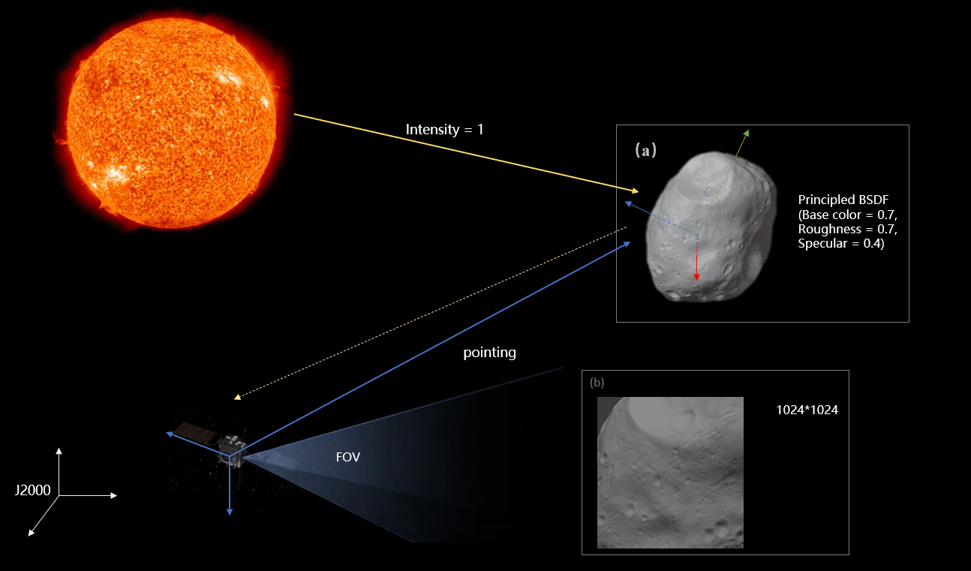}
    \caption{Illustration of the synthetic-image generation setup. The rendering is defined by the Sun--spacecraft--target geometry, the camera pointing and field of view, and the 3D shape model of \textit{Phobos}. Panel (a) shows the rendered shape-model view under the adopted illumination and material settings, and panel (b) shows the corresponding synthetic image.}  
    \label{fig:s_o_p}
\end{figure}

The synthetic image was rendered in Blender \citep{Blender}. Its purpose was not to reproduce the observed photometry in a fully physical sense, but to provide a reference image that is consistent with the observation geometry for establishing correspondences and for the subsequent geometric refinement. The rendering was specified by (i) the geometry obtained from SPICE, (ii) the adopted SRC projection model and the corrected pointing, and (iii) a simplified reflectance model used to generate surface shading for feature matching. Figure~\ref{fig:s_o_p} illustrates the procedure used to generate the synthetic image, including the geometric relationship among the Sun, the spacecraft, and the target, the view of \textit{Phobos} rendered from the shape model \citep{Ernst2023}, and the corresponding synthetic image.

\subsection{Feature matching}
\label{sec:matching}
We establish correspondences between the observed image $I_{\rm obs}$ and the
corresponding synthetic image $I_{\rm syn}$ using a learned feature-matching
pipeline comprising SuperPoint \citep{DeTone2018} and SuperGlue
\citep{Sarlin2020}. Despite differences in local photometric appearance, the
two images contain common geometric structures associated with the same
surface features, which can be exploited through local-feature matching
followed by geometric verification.

No additional photometric preprocessing is applied to reduce the
differences in intensity distribution between $I_{\rm obs}$ and $I_{\rm syn}$.
In particular, neither image is subjected to histogram matching, histogram
equalization, or other histogram-based intensity transformations, and
$I_{\rm obs}$ is not high-pass filtered or contrast enhanced. The
radiometrically calibrated $I_{\rm obs}$ and the rendered $I_{\rm syn}$ are
supplied directly to the matching pipeline.

\begin{figure}[h]
    \centering
    \includegraphics[width=1\linewidth]{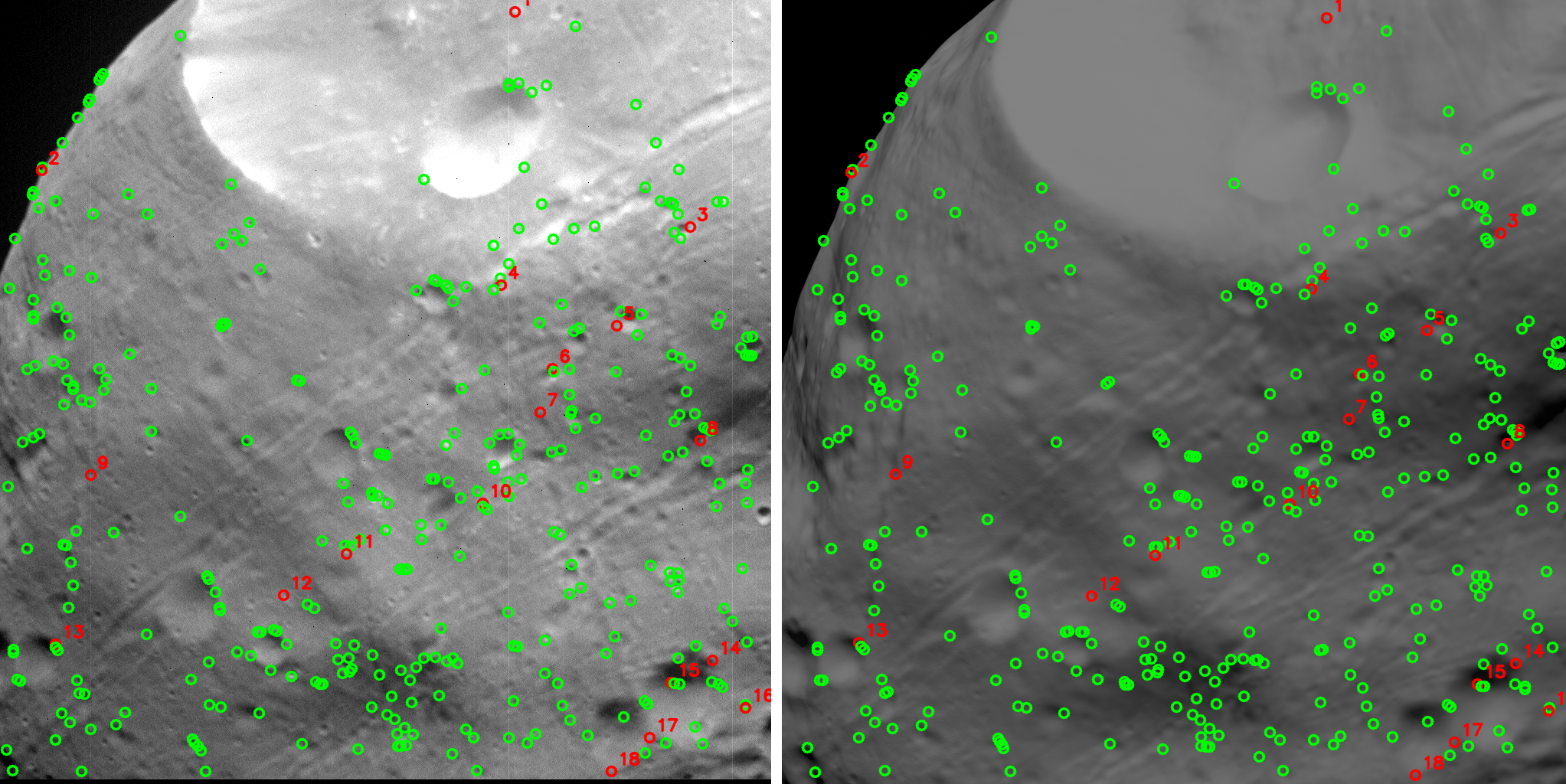}
    \caption{Example of the inlier correspondences retained after SuperPoint–SuperGlue matching and RANSAC filtering between an observed image (left) and its corresponding synthetic image (right). All retained keypoints are marked by green circles. For clarity,
    a subset of the matched pairs is highlighted by red circles and identical
    numerical labels in the two images. }    
    \label{fig:s_o_s_r}
\end{figure}

SuperPoint is applied to both images to detect keypoints and compute their
local descriptors. SuperGlue then generates tentative correspondences between
the keypoints in the two images. To reject mismatches and enforce geometric
consistency, we apply Random Sample Consensus (RANSAC) filtering \citep{Fischler1981}.
The resulting set of $N$ inlier correspondences is written as
\[
\{(\mathbf{u}^{\rm obs}_i,\mathbf{u}^{\rm syn}_i)\}_{i=1}^{N},
\]
where $\mathbf{u}^{\rm obs}_i$ and $\mathbf{u}^{\rm syn}_i$ are the
image-plane coordinates of the two keypoints forming the $i$th correspondence
in $I_{\rm obs}$ and $I_{\rm syn}$, respectively. An example is shown in
Figure~\ref{fig:s_o_s_r}.

\subsection{2D--3D association}
\label{sec:ray}
The inlier matches provide 2D--2D correspondences between $I_{\rm obs}$ and $I_{\rm syn}$. To obtain the 2D--3D constraints required for geometric adjustment, a viewing ray is constructed for each synthetic-image keypoint $\mathbf{u}^{\rm syn}_i$ by back-projecting its image coordinates through the camera center using
the same projection geometry employed to render $I_{\rm syn}$. The nearest valid intersection of this ray with the shape model is taken
as the associated surface point $\mathbf{X}_i$. Matches for which no valid intersection is found are discarded. The remaining correspondences form the set
\[
\{(\mathbf{u}^{\rm obs}_i,\mathbf{X}_i)\}_{i=1}^{M}.
\]

The association between $\mathbf{u}^{\rm obs}_i$ and $\mathbf{X}_i$ is approximate because the matched image features may not localize
exactly the same physical surface point. Nevertheless, the synthetic image is generated from a nominal geometry that is already close to
the observation, the retained matches are filtered for geometric consistency, and the subsequent adjustment estimates only small
corrections. Under these conditions, the resulting 2D--3D constraints provide a sufficiently stable basis for plane-of-sky position refinement.

\subsection{Geometric adjustment}
\label{sec:opt}

The initial geometry does not in general bring the synthetic rendering into exact agreement with the observed image. Residuals may arise from errors in the nominal position of \textit{Phobos}, the spacecraft position, the nominal body orientation, and the camera pointing, as well as from matching uncertainty and imperfections in the adopted shape model or rendering assumptions. Focusing on the target geometry, a general adjustment would include three rotational corrections to the body orientation and three translational corrections to the position of \textit{Phobos} relative to Mars Express. For a single image, however, these six parameters are strongly coupled and cannot be robustly and independently estimated without additional constraints. Increasing the number of feature matches reduces random matching uncertainty but does not remove this parameter degeneracy.
Because our objective is astrometry rather than rotational-state estimation, we hold the adopted body orientation fixed and adjust only the relative position of \textit{Phobos}. Of its three translational components, the component along the nominal observer-to-\textit{Phobos} line of sight is weakly constrained because it primarily affects the apparent scale and perspective of the resolved body. We therefore hold this line-of-sight component fixed, and estimate only two transverse position offsets in the plane of the sky. These offsets are effective corrections: they may absorb errors in the fixed geometric quantities and are therefore not interpreted as independent physical corrections to the Phobos ephemeris.
Let $\mathbf{r}_0$ denote the nominal vector from the observer to the COF of \textit{Phobos} in the J2000 inertial frame, and let
\begin{equation}
\hat{\mathbf{n}}
=
\frac{\mathbf{r}_0}{\lVert\mathbf{r}_0\rVert}
\end{equation}
be the corresponding unit vector. The transverse correction to the observer-to-COF vector is parameterized as
\begin{equation}
\Delta\mathbf{r}
=
\Delta x\,\hat{\mathbf{e}}_x
+
\Delta y\,\hat{\mathbf{e}}_y ,
\label{eq:sky_plane_correction}
\end{equation}
where the orthonormal plane-of-sky basis vectors are defined by
\begin{equation}
\hat{\mathbf{e}}_x
=
\frac{
\hat{\mathbf{c}}_s
-
(\hat{\mathbf{c}}_s\mathbin{\cdot}\hat{\mathbf{n}})
\hat{\mathbf{n}}
}{
\left\|
\hat{\mathbf{c}}_s
-
(\hat{\mathbf{c}}_s\mathbin{\cdot}\hat{\mathbf{n}})
\hat{\mathbf{n}}
\right\|
},
\qquad
\hat{\mathbf{e}}_y
=
\hat{\mathbf{n}}\times\hat{\mathbf{e}}_x .
\label{eq:sky_plane_basis}
\end{equation}
Here, $\hat{\mathbf{c}}_s$ is the unit vector along the positive camera sample direction, expressed in the J2000 frame. Under the adopted camera-coordinate convention, $\hat{\mathbf{e}}_x$ and $\hat{\mathbf{e}}_y$ are oriented approximately along the image sample and line directions, respectively. The offsets $\Delta x$ and $\Delta y$ are expressed in kilometres. By construction,
$\Delta\mathbf{r}\mathbin{\cdot}\hat{\mathbf{n}}=0$, so no correction is estimated along the nominal line of sight.
Let $\mathbf{X}_i$ be the position of the $i$th matched surface point relative to the COF of \textit{Phobos}, expressed in the body-fixed frame, and let $\mathbf{R}_0$ be the adopted rotation matrix from the body-fixed frame to the J2000 frame. With the body orientation fixed, the corrected vector from observer to surface point is
\begin{equation}
\boldsymbol{\rho}_i^{\mathrm{I}}
=
\mathbf{r}_0
+
\Delta \mathbf{r}
+
\mathbf{R}_0 \mathbf{X}_i .
\label{eq:inertial_los_twoparam}
\end{equation}
Here, $\mathbf{R}_0 \mathbf{X}_i$ gives the vector from the COF to the surface point after transforming it from the frame fixed to \textit{Phobos} to the inertial frame. The vector $\boldsymbol{\rho}_i^{\mathrm{I}}$ is then transformed into the camera frame,
\begin{equation}
\boldsymbol{\rho}_i^{\mathrm{c}}
=
\mathbf{A}\boldsymbol{\rho}_i^{\mathrm{I}},
\label{eq:camera_los_twoparam}
\end{equation}

where $\mathbf{A}$ is the rotation matrix from the inertial frame to the camera frame after applying the sequence-level pointing correction. The corresponding image point is obtained through the adopted SRC projection model,
\begin{equation}
\mathbf{u}_i
=
\boldsymbol{\pi}
\left(
\boldsymbol{\rho}_i^{\mathrm{c}}
\right),
\label{eq:image_projection_twoparam}
\end{equation}
where $\boldsymbol{\pi}(\cdot)$ denotes the projection from the camera frame to the image plane.

Given the $M$ retained 2D--3D correspondences
$\{(\mathbf{u}_i^{\rm obs},\mathbf{X}_i)\}_{i=1}^{M}$, the transverse
offsets are estimated by minimizing the total reprojection error:
\begin{equation}
\begin{aligned}
\min_{\Delta x,\,\Delta y}\quad
&
\sum_{i=1}^{M}
\left\|
\mathbf{u}_i^{\rm obs}
-
\boldsymbol{\pi}
\left(
\mathbf{A}
\left[
\mathbf{r}_0
+
\Delta x\,\hat{\mathbf{e}}_x
+
\Delta y\,\hat{\mathbf{e}}_y
+
\mathbf{R}_0\mathbf{X}_i
\right]
\right)
\right\|^2 .
\end{aligned}
\label{eq:twoparam_optimization}
\end{equation}

The forward model retains the exact perspective projection.
The resulting nonlinear optimization problem is solved using
the L-BFGS-B algorithm \citep{Byrd1995}, initialized with
$\Delta x=\Delta y=0$.

The optimized observer-to-COF vector in the Mars Express-centered J2000 frame is
$ \mathbf{r}=\mathbf{r}_0+\Delta\mathbf{r}^{*}$,
 where $\Delta\mathbf{r}^{*}$ is the optimized plane-of-sky correction.
The corresponding right ascension and declination are
\begin{equation}
    \alpha
    = \operatorname{atan2}(r_y,r_x),
    \qquad
    \delta
    = \arcsin\!\left(\frac{r_z}{\lVert\mathbf{r}\rVert}\right),
\end{equation}
where $(r_x,r_y,r_z)$ are the Cartesian components of $\mathbf{r}$, negative values returned by $\operatorname{atan2}$ are wrapped
into the interval $[0,2\pi)$.

\subsection{Quality control and outputs}
\label{sec:qc}
Not all images yield enough reliable correspondences for stable estimation. To ensure the overall reliability of the matched features, we first apply RANSAC to perform initial geometric filtering. Subsequently, a Median Absolute Deviation (MAD) based filter is utilized to robustly eliminate any remaining spatial outliers. An image is accepted only if at least six correspondences remain after all rejection steps. Since solutions based on very few correspondences are often weakly constrained and numerically unstable in practice, this threshold serves as a conservative empirical minimum for solution acceptance. When applied to the extended SRC dataset analysed in
Section~\ref{sec:extended}, the pipeline rejected 47 of the 1160 input
images because they did not yield acceptable astrometric solutions under
the above quality-control procedure. The remaining 1113 images were
retained for the final astrometric analysis.

For each successfully processed image, the pipeline outputs optimized geometric corrections, the derived COF position and corresponding astrometric measurement, and quality indicators such as the number of inlier correspondences.

\section{Data and ancillary products}
\label{sec:valdata}

We used standard radiometrically calibrated SRC products from the following PDS3 datasets: MEX-M-HRSC-3-RDR-EXT2-V4, MEX-M-HRSC-3-RDR-EXT3-V4, MEX-M-HRSC-3-RDR-EXT4-V4, MEX-M-HRSC-3-RDR-EXT5-V4, MEX-M-HRSC-3-RDR-EXT6-V4, MEX-M-HRSC-3-RDR-EXT7-V4, MEX-M-HRSC-3-RDR-EXT8-V4, and MEX-M-HRSC-3-RDR-EXT9-V4 \citep{Roatsch}. The data were retrieved from the PDS Geosciences Node (\url{https://pds-geosciences.wustl.edu/mex}). After the standard operational cropping, the images contain 1008 $\times$ 1018 pixels, with a small number of images having 1113 rows. The images used in this work were selected according to two practical requirements: each observing sequence must include the long-exposure frames required for the sequence-level pointing correction applied before the astrometric reduction, and the \textit{Phobos} disk must be sufficiently resolved to provide usable surface texture for automated feature detection and matching.

\begin{table}[htbp]
\centering
\begin{threeparttable}

\caption{SRC geometric camera-model and frame-alignment parameters
adopted in this work.}
\label{tab:SRC_Calibration}

\begin{tabular}{@{}c@{}}
\hspace*{-1.6cm}\begin{tabular}{@{}cccccc@{}}
\toprule
\multicolumn{6}{c}{\textbf{Camera-model parameters}} \\
\cmidrule(lr){1-6}
$F_a$ (mm) &
$1/K$ (mm\,px$^{-1}$) &
$S_0$ (px) &
$L_0$ (px) &
$\alpha_1$ (px\,mm$^{-3}$) &
$\alpha_2$ (px\,mm$^{-5}$) \\
\midrule
984.76 &
0.009 &
512.5 &
512.5 &
$-6.644\times10^{-5}$ &
$3.087\times10^{-6}$ \\
\midrule
\multicolumn{6}{c}{\textbf{SRC--HRSC frame-alignment angles}} \\
\cmidrule(lr){1-6}
\multicolumn{2}{c}{$\theta_X$ (deg)} &
\multicolumn{2}{c}{$\theta_Y$ (deg)} &
\multicolumn{2}{c}{$\theta_Z$ (deg)} \\
\midrule
\multicolumn{2}{c}{$-0.084154$} &
\multicolumn{2}{c}{$-0.038531$} &
\multicolumn{2}{c}{$90.038$} \\
\bottomrule
\end{tabular}\hspace*{1.6cm}
\end{tabular}

\begin{tablenotes}[flushleft]
\footnotesize
\item \textit{Note.}
Here $F_a$ is the effective focal length, $1/K$ is the pixel
size, $S_0$ and $L_0$ define the principal point in the SRC full-frame
detector coordinate system, $\alpha_1$ and $\alpha_2$ are radial-distortion coefficients, and $\theta_X$, $\theta_Y$, and $\theta_Z$ are the SRC--HRSC frame-alignment angles.
\end{tablenotes}

\end{threeparttable}
\end{table}

In addition to the SRC images, the pipeline requires ancillary data defining the camera model, nominal observation geometry, and three-dimensional shape of \textit{Phobos}. We adopt the SRC camera-model and frame-alignment parameters listed in Table~\ref{tab:SRC_Calibration}. The intrinsic parameters and radial-distortion coefficients define a pinhole projection with image-plane radial distortion \citep{Pasewaldt2015}, while the alignment angles define the transformation between the SRC and HRSC reference frames. The principal point, $(S_0,L_0)=(512.5,512.5)$, is specified in the $1024\times1024$ full-frame detector coordinate system. Because the observed SRC products are cropped to a nominal size of $1008\times1018$ pixels, the coordinates of the matched observed-image keypoints are converted to full-frame coordinates using the corresponding crop offsets before ray back-projection and geometric adjustment. The same camera model is used throughout synthetic-image rendering, ray back-projection, and reprojection.

The nominal observation geometry is computed using SPICE kernels for Mars Express, \textit{Phobos}, and the relevant reference frames. These kernels provide the spacecraft trajectory, camera attitude, body orientation, and ephemeris information required to compute the viewing and illumination geometry. The resulting geometry is used both to render the initial synthetic images and to initialize the geometric adjustment. The SPICE meta-kernel used in this study, which lists the required kernel files, is available in the accompanying repository.\footnote{\url{https://github.com/Astrometry-JNU/Feature-Matching-Data/blob/main/spice_kernel.tm}}

We use the high-resolution shape model of \textit{Phobos} developed by \citet{Ernst2023} from multi-mission images using stereophotoclinometry. The model provides global coverage, a mean facet scale of approximately 18~m, vertex spacing down to $\sim$10~m in the best-constrained regions, and local surface precision better than $\sim$5~m. These properties
make it suitable for the geometric operations in the astrometric pipeline.

\section{Astrometric results}
\label{sec:results}
To assess the performance of the automated pipeline, we first reprocess the 130 SRC images analysed by \citet{Pasewaldt2015} for direct comparison with the published control-point solution, and then apply the pipeline to an extended SRC dataset spanning 2007--2025. The resulting astrometric measurements derived from the extended dataset are evaluated against two independent reference ephemerides: the JPL Phobos ephemeris MAR099 \citep{Brozovic2025} and the NOE-4-2020 ephemeris \citep{Lainey2021}. For each exposure, the observed value (O) is the image-derived apparent direction to the Phobos COF, whereas the computed value (C) is the apparent direction predicted by the reference ephemeris at the exposure midtime. Both are computed with the SPICE \texttt{LT+S} aberration correction consistent with Sect.~\ref{sec:render}, and expressed in the Mars Express-centered J2000 frame. The O$-$C residuals are reported in the local tangent-plane components $\Delta\alpha\cos\delta$ and $\Delta\delta$, and are converted to kilometres using the Mars Express--Phobos distance at the exposure midtime.

\subsection{Comparison with the published control-point solution}
\label{sec:repro}
To compare the automated pipeline directly with the published control-point solution, we reprocessed the same 130 SRC images analysed by \citet{Pasewaldt2015}. The automated pipeline returned valid solutions for 127 images; the remaining three were rejected because sufficiently reliable feature correspondences could not be established. The comparison was therefore restricted to   this common 127-image subset, and the O$-$C residuals of both solutions were computed with respect to the same reference ephemeris, namely the JPL \texttt{MAR085} solution \citep{Jacobson2010}. 

The O$-$C residual statistics are summarized in Table~\ref{tab:oc_repro}. For a consistent comparison, the statistics for \citet{Pasewaldt2015} were recomputed from their
published measurements for the common 127-image subset, rather than taken from their original 130-image statistics. The two solutions show nearly identical residual scatter: the automated pipeline has a slightly larger standard deviation in $\Delta\alpha\cos\delta$ (0.501~km versus 0.488~km) and a slightly smaller standard deviation in $\Delta\delta$ (0.900~km versus
0.906~km). Their mean residuals differ by 0.142~km and 0.035~km in the two components, respectively, which is small relative to the
corresponding residual scatter. The automated pipeline therefore achieves astrometric performance comparable to that of the published
control-point solution on the common dataset.

\begin{table}[!htbp]
\caption{O$-$C residual statistics for the published control-point solution and the present automated solution over the common 127-image subset. All residuals are computed with respect to the JPL \texttt{MAR085} ephemeris.}
\label{tab:oc_repro}
\centering
\begin{tabular}{lcccccc}
\toprule
Method & Component & Unit & $Min$ & $Max$ & $Mean$ & $Std$ \\
\midrule
Pasewaldt et al. (2015) & $\Delta\alpha\cos\delta$ & km & $-2.082$ & $0.853$ & $-0.291$ & $0.488$ \\                                           & $\Delta\delta$           & km & $-2.553$ & $4.102$ & $ 0.041$ & $0.906$ \\
\midrule
This work               & $\Delta\alpha\cos\delta$ & km & $-2.025$ & $0.838$ & $-0.149$ & $0.501$ \\                                           & $\Delta\delta$           & km & $-2.500$ & $4.321$ & $0.076$ & $0.900$ \\
\bottomrule
\end{tabular}\hspace{1.6cm}
\end{table}

\subsection{Extended dataset: catalog and residual statistics}
\label{sec:extended}

The automated pipeline was applied to 1160 SRC images acquired between 2007 and 2025. Of these, 47 images were rejected by the quality-control procedure described in Section~\ref{sec:qc}, leaving 1113 images with valid astrometric solutions. For each successfully processed image, the catalog includes the image identifier, the mid-exposure UTC, the derived astrometric position in right ascension and declination, the number of matched feature points retained in the final solution, and the corresponding Cartesian position vector expressed in the Mars Express-centered J2000 frame. A representative sample is listed in Table~\ref{tab:mar099_1113}, and the complete catalog is available online at \url{https://github.com/Astrometry-JNU/Feature-Matching-Data}.

\begin{figure}[htbp]
    \centering
    \includegraphics[width=0.90\linewidth]{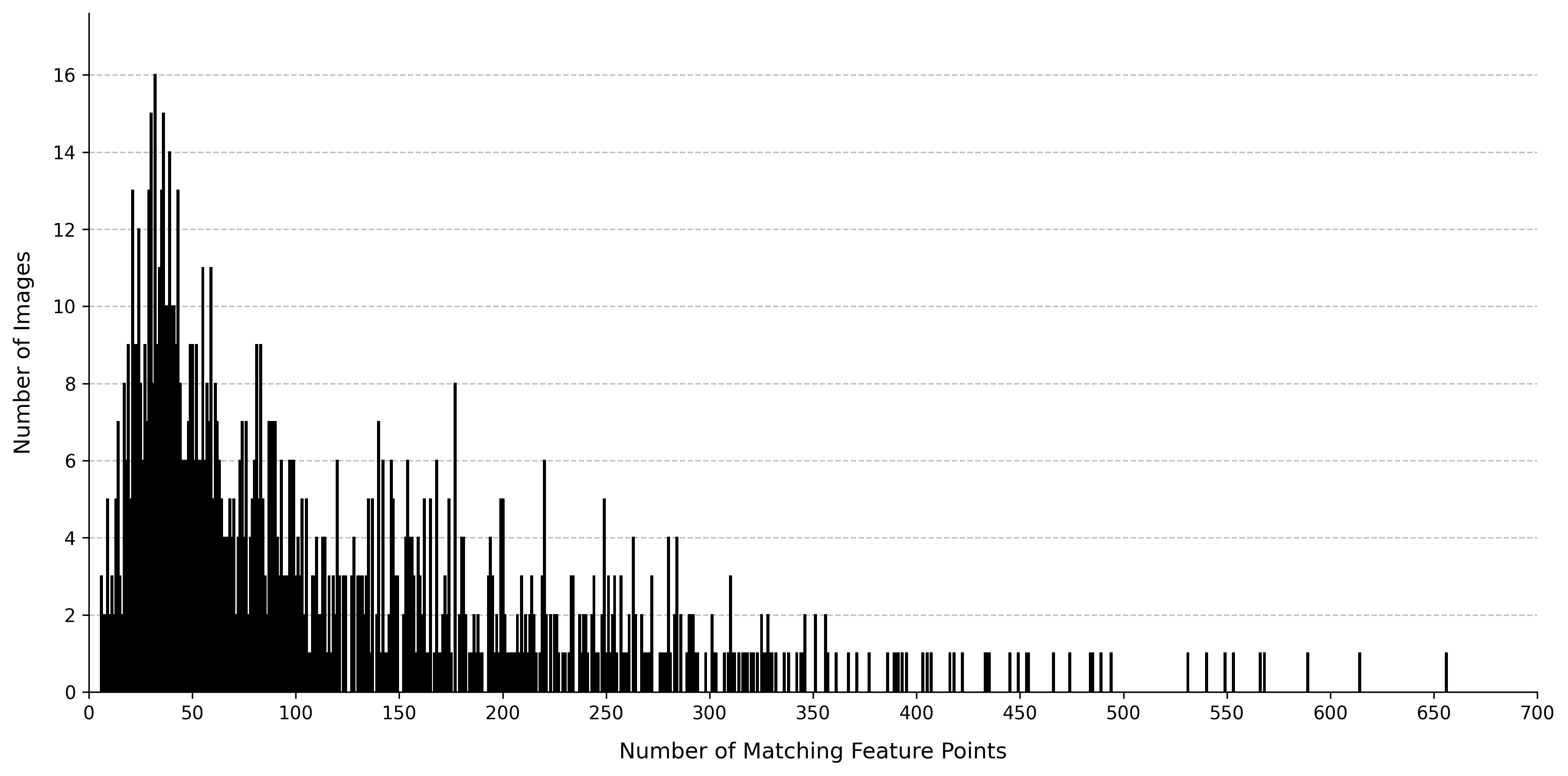}
    \caption{Distribution of the number of feature matches retained in the final astrometric solutions for the 1113-image dataset.}
    \label{fig:h_o_f_m}
\end{figure}

\begin{table}[!htbp]
  \centering
  \caption{A sample of the 1113 astrometric solutions. The column Method identifies the reduction method; FM denotes the feature-matching astrometric pipeline developed in this work. The columns $\alpha$ and $\delta$ give the apparent right ascension and declination of the Phobos COF as seen from Mars Express, referred to the J2000 equatorial axes. Column $N$ gives the number of matched feature points retained in the final solution. The columns $X$, $Y$, and $Z$ give the corresponding Cartesian position vector of the Phobos COF in the Mars Express-centered J2000 frame, in kilometres.}
  \label{tab:mar099_1113}
  \begin{tabular}{l l c r r r r r r}
    \toprule
    Image & UTC (mid-exp.) & Method & $\alpha$ (deg) & $\delta$ (deg) & $N$ & $X$ (km) & $Y$ (km) & $Z$ (km) \\
    \midrule
    HB911\_0003 & 2013-05-14 04:08:01.741 & FM & 11.4105 & 15.9193 & 91  & $3337.62$ & $673.62$   & $971.16$   \\
    HB911\_0004 & 2013-05-14 04:08:05.556 & FM & 11.6195 & 15.8653 & 251 & $3332.01$ & $685.15$   & $966.78$   \\
    HB911\_0005 & 2013-05-14 04:08:09.371 & FM & 11.8290 & 15.8122 & 239 & $3326.39$ & $696.68$   & $962.48$   \\
    HB911\_0006 & 2013-05-14 04:08:13.186 & FM & 12.0400 & 15.7620 & 58  & $3320.71$ & $708.26$   & $958.37$   \\
    HB940\_0004 & 2013-05-22 16:13:32.849 & FM & 320.0734 & $-$30.7064 & 93  & $5223.93$ & $-4372.01$ & $-4045.72$ \\
    HB940\_0005 & 2013-05-22 16:13:39.389 & FM & 320.2144 & $-$30.5981 & 99  & $5248.93$ & $-4371.00$ & $-4039.29$ \\
    HB940\_0006 & 2013-05-22 16:13:45.930 & FM & 320.3515 & $-$30.4871 & 97  & $5273.82$ & $-4370.41$ & $-4032.51$ \\
    HB963\_0004 & 2013-05-29 09:00:42.145 & FM & 324.9724 & $-$30.5001 & 97  & $5915.68$ & $-4146.45$ & $-4255.36$ \\
    HB963\_0005 & 2013-05-29 09:00:49.230 & FM & 325.1098 & $-$30.3925 & 99  & $5942.49$ & $-4144.03$ & $-4249.18$ \\
    HB963\_0006 & 2013-05-29 09:00:56.315 & FM & 325.2438 & $-$30.2832 & 93  & $5969.20$ & $-4141.93$ & $-4242.74$ \\
    HB992\_0004 & 2013-06-06 18:21:07.276 & FM & 8.9014  & $-$5.3837  & 133 & $4198.08$ & $657.51$   & $-400.45$  \\
    HB992\_0005 & 2013-06-06 18:21:11.637 & FM & 9.1081  & $-$5.4827  & 207 & $4192.31$ & $672.11$   & $-407.54$  \\
    HB992\_0006 & 2013-06-06 18:21:15.997 & FM & 9.3208  & $-$5.5705  & 134 & $4186.57$ & $687.14$   & $-413.79$  \\
    \bottomrule
  \end{tabular}
\end{table}
Across the 1113 solutions, the number of feature correspondences retained for the final adjustment ranges from the
acceptance threshold of 6 to 656, with a mean of 120 and a standard deviation of 106 (Table~\ref{tab:feature_stats}). The distribution is strongly right-skewed (Figure~\ref{fig:h_o_f_m}), with correspondence
counts concentrated below approximately 150 and a sparse tail extending to 656. This substantial image-to-image variation in the number of available geometric constraints likely reflects differences in image quality, viewing and illumination geometry, and visible surface texture.

\begin{table}[htbp]
\centering
\caption{Statistics of the number of retained feature matches per image across the 1113-image dataset.}
\label{tab:feature_stats}
\begin{tabular}{cccc}
\toprule
Mean &Minimum & Maximum  & Standard deviation ($\sigma$) \\
\midrule
120 & 6 & 656  & 106 \\
\bottomrule
\end{tabular}\hspace{1.6cm}
\end{table}

\begin{figure}[!htbp]
    \centering
    \includegraphics[width=0.90\linewidth]{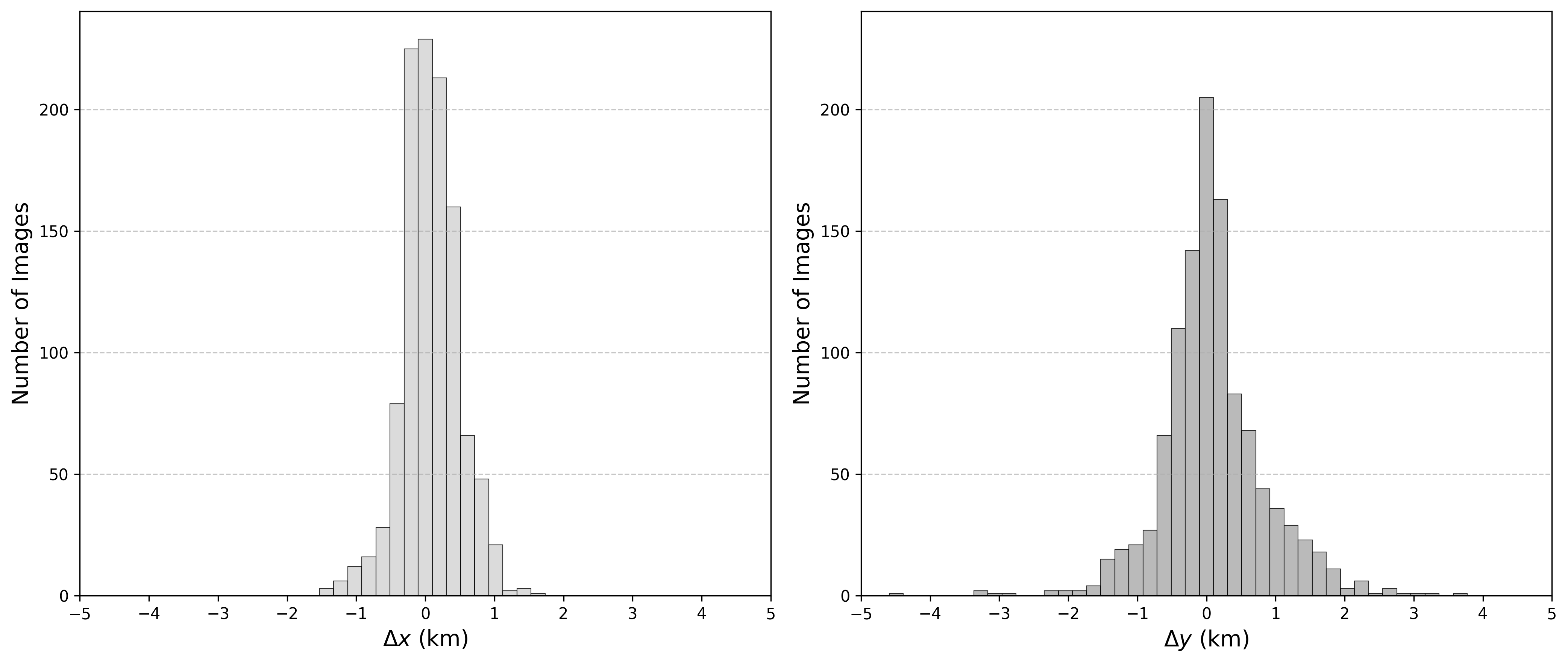}
    \caption{Distributions of the effective plane-of-sky offsets $\Delta x$ (left) and $\Delta y$ (right) for the 1113 valid astrometric solutions.}
    \label{fig:correction_hist}
\end{figure}

Each solution yields a pair of effective plane-of-sky corrections, $\Delta x$ and $\Delta y$, to the nominal position of \textit{Phobos}. As shown in Table~\ref{tab:correction_stats} and
Figure~\ref{fig:correction_hist}, both distributions are concentrated near zero, and the corrections are generally of order 1~km or less.
This is consistent with the assumption that the nominal geometry is already close to the observation and requires only a small local
adjustment. The larger dispersion in $\Delta y$ indicates greater image-to-image variation in this component.

\begin{table}[!htbp]
    \centering
    \caption{Statistics of the effective plane-of-sky offsets $\Delta x$ and $\Delta y$ for the 1113 valid astrometric solutions.}
    \label{tab:correction_stats}
    \begin{tabular}{cccccc}
        \toprule
         Component & Unit & $Min$ & $Max$ & $Mean$ & $Std$ \\
        \midrule                 
           $\Delta x$ & km & $-1.360$ & $1.596$ & $0.077$ & $0.411$ \\   
           $\Delta y$ & km & $-4.508$ & $3.596$ & $0.093$ & $0.751$ \\   
        \bottomrule
    \end{tabular}\hspace{1.6cm}
\end{table}

To evaluate the astrometric behaviour of the extended dataset, we computed O$-$C residuals for the 1113 valid measurements with respect to the JPL MAR099 and NOE-4-2020 ephemerides. The full residual statistics are listed in Table~\ref{tab:oc_1113}. Relative to MAR099, the mean residuals are 0.186~km in $\Delta\alpha\cos\delta$ and 0.053~km in $\Delta\delta$, with corresponding standard deviations of 0.609 and 0.583~km. The NOE-4-2020 comparison gives slightly larger standard deviations of 0.670 and 0.651~km, respectively. For both ephemerides, the absolute mean residuals remain below 0.3~km, while the residual scatter is modestly smaller relative to MAR099 in both components.

\begin{figure}[!htbp]
    \centering
    \includegraphics[width=0.7\linewidth]{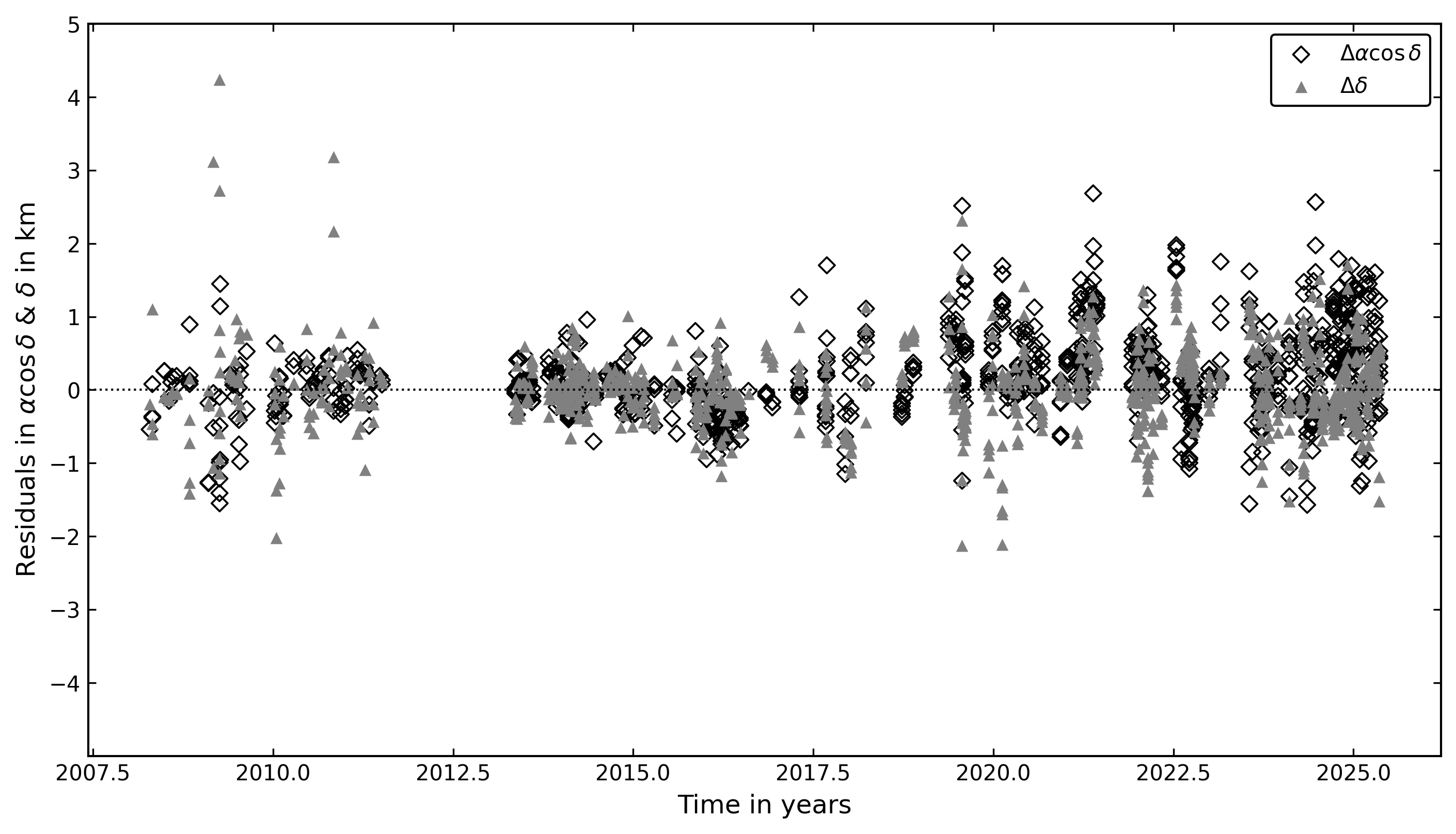}
    \caption{O$-$C residuals of the full dataset as a function of time with respect to the JPL MAR099 ephemeris.}
    \label{fig:oc_jpl}
\end{figure}

\begin{figure}[!htbp]
    \centering
    \includegraphics[width=0.7\linewidth]{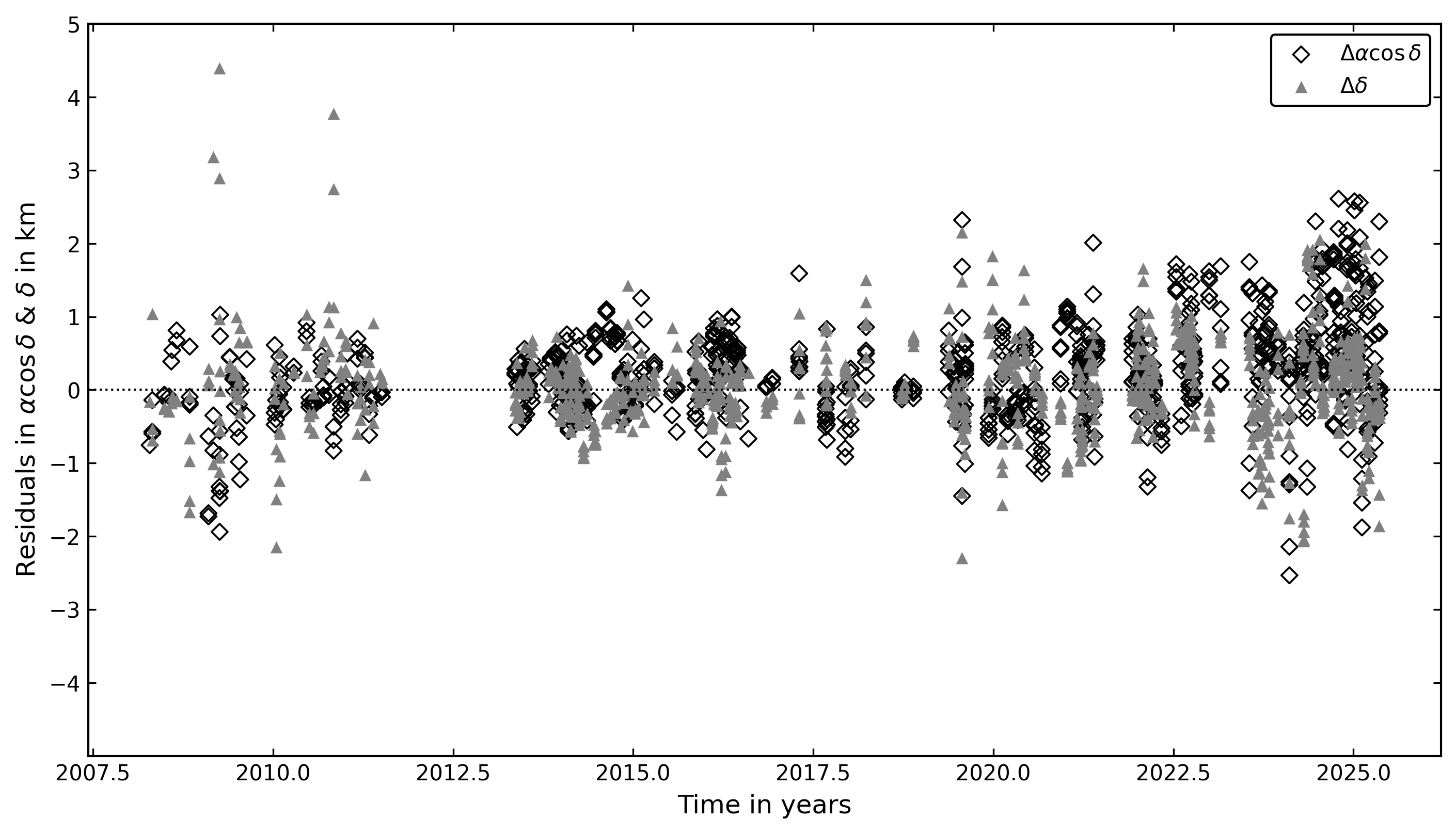}
    \caption{O$-$C residuals of the full dataset as a function of time with respect to the NOE-4-2020 ephemeris.}
    \label{fig:oc_noe}
\end{figure}

Figures~\ref{fig:oc_jpl} and~\ref{fig:oc_noe} show the O$-$C residuals as a function of time for the two reference ephemerides. Over the full 2007--2025 interval, no obvious long-term drift or strong time-dependent structure is apparent in either $\alpha\cos\delta$ or $\delta$.

\begin{table}[!htbp]
    \centering
    \caption{O$-$C residual statistics for the 1113-image dataset with respect to the JPL MAR099 and NOE-4-2020 ephemerides.}
    \label{tab:oc_1113}
    \begin{tabular}{lcccccc}
        \toprule
        Ephemeris & Component & Unit & $Min$ & $Max$ & $Mean$ & $Std$ \\
        \midrule        
        \multirow{2}{*}{MAR099}          
          & $\Delta\alpha\cos\delta$ & km & $-2.211$ & $2.772$ & $ 0.186$ & $0.609$ \\          
          & $\Delta\delta$           & km & $-2.176$ & $4.239$ & $ 0.053$ & $0.583$ \\        
        \midrule    
        \multirow{2}{*}{NOE-4-2020}          
          & $\Delta\alpha\cos\delta$ & km & $-2.533$ & $2.772$ & $ 0.249$ & $0.670$ \\          
          & $\Delta\delta$           & km & $-2.293$ & $4.391$ & $ 0.068$ & $0.651$ \\        
        \bottomrule
    \end{tabular}\hspace{1.6cm}
\end{table}

\section{Discussion}
\label{sec:discussion}
On the comparison dataset, the automated pipeline yields O$-$C residual scatter comparable to that of the published control-point reduction, demonstrating comparable astrometric performance. Unlike classical control-point reductions, which may depend on operator choices in landmark selection and correspondence validation, the
pipeline applies uniform feature-matching, outlier-rejection, and geometric-adjustment procedures to all images. Across the
extended dataset, it retains an average of 120 correspondences per accepted image, providing redundancy against individual mismatches and supporting stable estimation of the plane-of-sky offsets.

For the extended dataset, the residual dispersions are modestly smaller with respect to JPL MAR099 than to NOE-4-2020, indicating that the measurements are slightly more consistent with MAR099. For both reference ephemerides, the residual dispersions remain at the sub-kilometre level over a long observational baseline.
Several practical limitations remain. The success of feature matching depends on the visibility of surface texture in the illuminated portion of the \textit{Phobos} disk. Low contrast, extensive shadowing, and highly oblique illumination can reduce the number and spatial distribution of reliable correspondences, introducing an observing-geometry-dependent selection effect into the final dataset. The measurements also depend on the fidelity of the shape model and forward rendering: local shape errors affect the ray--shape association, while photometric differences between observed and synthetic images may produce geometry-dependent biases.
A further limitation is that the fitted offsets do not have a unique physical interpretation. The single-image geometry does not independently constrain the body orientation and all three components of the observer-to-\textit{Phobos} position. Consequently, the fitted transverse offsets may absorb errors in the fixed body orientation, spacecraft state, and camera pointing and should be regarded as effective astrometric corrections rather than as independent corrections to the Phobos ephemeris.

These limitations motivate several directions for future development, including improved camera and shape models, more realistic synthetic rendering, and more robust feature matching under low-contrast or challenging illumination conditions.
With appropriate measurement weighting and treatment of geometry-dependent systematic effects, the uniformly reduced 2007--2025 dataset provides a useful basis for Phobos ephemeris refinement and dynamical analyses. More broadly, the results demonstrate the potential of automated synthetic-to-observed matching for large-scale, homogeneous astrometric reduction of disk-resolved spacecraft images.

\section{Conclusions}
\label{sec:conclusion}
We present an automated shape-model-based pipeline for astrometry of \textit{Phobos} in Mars Express SRC images. It combines synthetic-image rendering, feature matching, ray--shape association, and a per-image geometric adjustment that estimates two effective plane-of-sky offsets while holding the adopted body orientation, spacecraft state, and corrected camera pointing fixed. Applied to 1160 selected images, the pipeline produced 1113 valid measurements spanning 2007--2025, forming a uniformly reduced astrometric dataset from the SRC archive.

On the comparison dataset, the automated solution yields O$-$C residual scatter comparable to that of the published control-point solution under the same JPL \texttt{MAR085} reference ephemeris. For the extended dataset, the residual scatter is at the sub-kilometre level in both sky-plane components relative to the JPL MAR099 and NOE-4-2020 ephemerides, with no obvious long-term trend.

Although performance remains sensitive to surface-feature visibility, observing geometry, and the fidelity of the shape model and synthetic rendering, the pipeline enables consistent reduction of a large spacecraft-image archive without manual landmark selection. The resulting dataset provides new astrometric input for Phobos ephemeris refinement and dynamical analyses, while the underlying framework offers a practical route to homogeneous astrometry of other disk-resolved bodies.

\begin{acknowledgments}
This work has been supported by the National Natural Science Foundation of China  (No. 12373073, No. 12461160265, No. U2031104) and the Guangdong Basic and Applied Basic Research Foundation (No. 2023A1515011340, No. 20241515011762). This work has made use of data from the European Space Agency (ESA) mission {\it Gaia} (\url{https://www.cosmos.esa.int/gaia}), processed by the {\it Gaia} Data Processing and Analysis Consortium (DPAC, \url{https://www.cosmos.esa.int/web/gaia/dpac/consortium}). Funding for the DPAC has been provided by national institutions, in particular the institutions participating in the {\it Gaia} Multilateral Agreement. 
\end{acknowledgments}


%
\facilities{MEX(HRSC, SRC)}

\software{PyTorch \citep{paszke2019pytorch},
          SuperGlue \citep{Sarlin2020},
          SpiceyPy \citep{annex2020spiceypy},
          Blender \citep{Blender}
          }


\bibliography{main}{}
\bibliographystyle{aasjournalv7}



\end{document}